\documentclass[%
 reprint,
 amsmath,amssymb,
 aps,
]{revtex4-2}

\usepackage{graphicx}% Include figure files
\usepackage{dcolumn}% Align table columns on decimal point
\usepackage{bm}% bold math
\usepackage{url}

\begin{document}

\preprint{APS/123-QED}

\title{Germanium response to sub-keV nuclear recoils:\\ a multipronged experimental characterization}% Force line breaks with \\
%\thanks{A footnote to the article title}%

\author{J.I. Collar}
\email{collar@uchicago.edu}
\author{A.R.L. Kavner}
\author{C.M. Lewis}%
\affiliation{%
Enrico Fermi Institute, Kavli Institute for Cosmological Physics, and Department of Physics\\
University of Chicago, Chicago, Illinois 60637, USA
}%

%\collaboration{CLEO Collaboration}%\noaffiliation

\date{\today}% It is always \today, today,
             %  but any date may be explicitly specified

\begin{abstract}
Germanium is the detector material of choice in many rare-event searches looking for low-energy nuclear recoils induced by dark matter particles or neutrinos. We perform a systematic exploration of its quenching factor for sub-keV  nuclear recoils, using multiple techniques: photo-neutron sources, recoils from gamma-emission following thermal neutron capture, and a monochromatic filtered neutron beam. Our results point to a marked deviation from the predictions of the Lindhard model in this mostly unexplored energy range. We comment on the compatibility of our data with low-energy processes such as the Migdal effect, and on the impact of our measurements on upcoming searches. 
\end{abstract}

%\keywords{Suggested keywords}%Use showkeys class option if keyword
                              %display desired
\maketitle

%\tableofcontents

\section{\label{sec:level1}Introduction}

The study of low-energy nuclear recoils (NRs) induced by the elastic scattering of neutral particles off nuclei is an active area of research in  particle physics. Until recently, its main motivation was the numerous experiments looking for Weakly Interacting Massive Particles (WIMPs), one of the most popular and well-motivated dark matter candidates. The recent experimental demonstration of Coherent Elastic Neutrino-Nucleus Scattering (CE$\nu$NS) \cite{freedman,science,ournim,bjorn,nicolethesis}, a process involving this same mechanism of interaction -albeit from particles known to exist- has reinforced the need to understand signal generation from these subtle NRs in a variety of  materials.

For detecting media exploiting the ionization or scintillation generated by particle interactions, a central quantity is the so-called quenching factor (QF). This is the ratio of observable energy expressed in one of  those channels by a NR, over that generated by an electron recoil (ER) of the same kinetic energy. In experimental studies, the first are typically generated by fast neutrons, the second by gammas or x-rays. At NR energies of interest (few keV$_{nr}$) this QF is typically of order 10\%, adding to the difficulty of WIMP and CE$\nu$NS searches. We have recently emphasized the importance of dedicated QF calibrations able to discern its energy dependence \cite{csiqf}. These studies are fundamental in order to access the many deviations from the Standard Model, involving new physics, that are testable via CE$\nu$NS. Experimental efforts shirking QF characterization are subject to  uncertainties in signal significance and interpretation \cite{lar,conus2}. 

Low-noise p-type point contact (PPC) detectors \cite{ppc} can register the ionization from sub-keV energy depositions in large ($>$1 kg) germanium crystals. As a result of this and other virtues, PPCs are presently used in searches for neutrinoless double-beta decay \cite{majorana,gerda,legend}, WIMPs \cite{cogent,cdex}, CE$\nu$NS \cite{ESS,conus1,texono,nugen}, and exotic modes of particle decay \cite{muon}. This work concentrates on the characterization of the QF for ionization-sensitive germanium detectors in the sub-keV$_{nr}$ NR energy region. This realm remains essentially unexplored for most materials. The information obtained strongly impacts future germanium searches for CE$\nu$NS from reactor antineutrinos  and for low-mass WIMPs.

In order to access the tiny energies involved in this study, down to $\sim$50 eV in deposited ionization, we employ a small (1 cm$^{3}$, 78 eV FWHM noise) GL0110 LEGe (Low Energy Germanium) detector \cite{lege}. For crystals this size the use of n-type germanium is possible while preserving the good charge collection and energy resolution seen to rapidly degrade for larger n-type point-contact configurations \cite{ppc,Luke}. This choice removes the nuisance parameters introduced in the analysis by the O(1) mm thickness of inert surface electrode layers in p-type material \cite{bjornqf,deadl}. At sub-micron thickness for this device, this surface structure can be safely neglected. In addition to this, continuous advancements in noise reduction for point-contact detectors \cite{canberra} allowed us to reach a 200 eV ionization energy analysis threshold in this device and even smaller ($\sim$100 eV) for externally-triggered signals. This is in contrast to the 1 keV  threshold achieved in our latest germanium QF study \cite{bjornqf}. A reduced noise also results in  excellent energy resolution (Fig.\ 1). Lastly, multiple scattering in this small LEGe involves just a \mbox{4\%-17\%} of interactions, depending on calibration technique. Multiple scatters dominated previous studies using $\sim$100 cm$^{3}$  PPCs \cite{phil,bjornqf}, limiting the  modes of analysis available.  

This paper is organized chronologically. It tells a story of experimentation spanning three years, where the unexpected results from a first QF study led to a total of four calibration techniques being used in an attempt to validate or refute those. The process involved a variety of neutron sources selected to populate the sub-keV$_{nr}$ region sought. The net outcome is a strong case for a sharply increasing ionization yield in germanium with decreasing NR energy below $\sim$1.3 keV$_{nr}$, in clear departure from the Lindhard model \cite{lindhard} typically assumed for this material. We conclude by briefly commenting on several physical processes able to lead to our observations and on their impact on upcoming searches for rare-events, while encouraging further QF characterization work by others.

\section{\label{sec:level1}photo-neutron sources: $^{88}$Y/Be}

Profiting from the factor of five improvement in energy threshold in the new LEGe compared to the larger PPC employed in \cite{bjornqf}, we revisited the technique implemented there, proposed in \cite{ybeprl} and  previously also used for silicon QF characterization \cite{alvaro}. Briefly reviewed, a photo-neutron $^{88}$Y/Be source generates monochromatic 152 keV neutrons from beryllium photo-disintegration accompanied by a much more intense high-energy gamma emission, which can nevertheless be blocked by 15-20 cm of lead while causing only a minimal degradation of neutron energies. Additional data taken with a $^{88}$Y/Al source configuration isolate any remaining events from an otherwise unchanged gamma component: for $^{88}$Y emissions, the gamma stopping of Al and BeO are equivalent, while no photo-disintegration is possible for Al. The residual spectrum from the difference of both runs contains  NR contributions only \cite{ybeprl}. 

For germanium, the maximum recoil energy produced by the elastic scattering of these neutrons is 8.5 keV$_{nr}$. As in \cite{bjornqf} a 5 mCi $^{88}$Y source \cite{procured} was evaporated into a triple-sealed container, placed within a BeO ceramic gamma-to-neutron converter. Over the one month of LEGe exposure to the source ($T_{1/2}$=106.6 d), its average neutron yield was 848 n/s. This was  measured indirectly via gamma spectroscopy followed by a calculation involving a revised $^{9}\mathrm{Be}(\ensuremath{\gamma},n)^{8}\mathrm{Be}$ cross section \cite{alan}, and directly with a  $^{3}$He counter surrounded by moderator. In this case, both methods agreed within $\sim$2\%. Our past direct measurements of neutron sources show disagreements in yield  with manufacturer data of up to 24\% \cite{drew}. 

\begin{figure}[!htbp]
\includegraphics[width=.93 \linewidth]{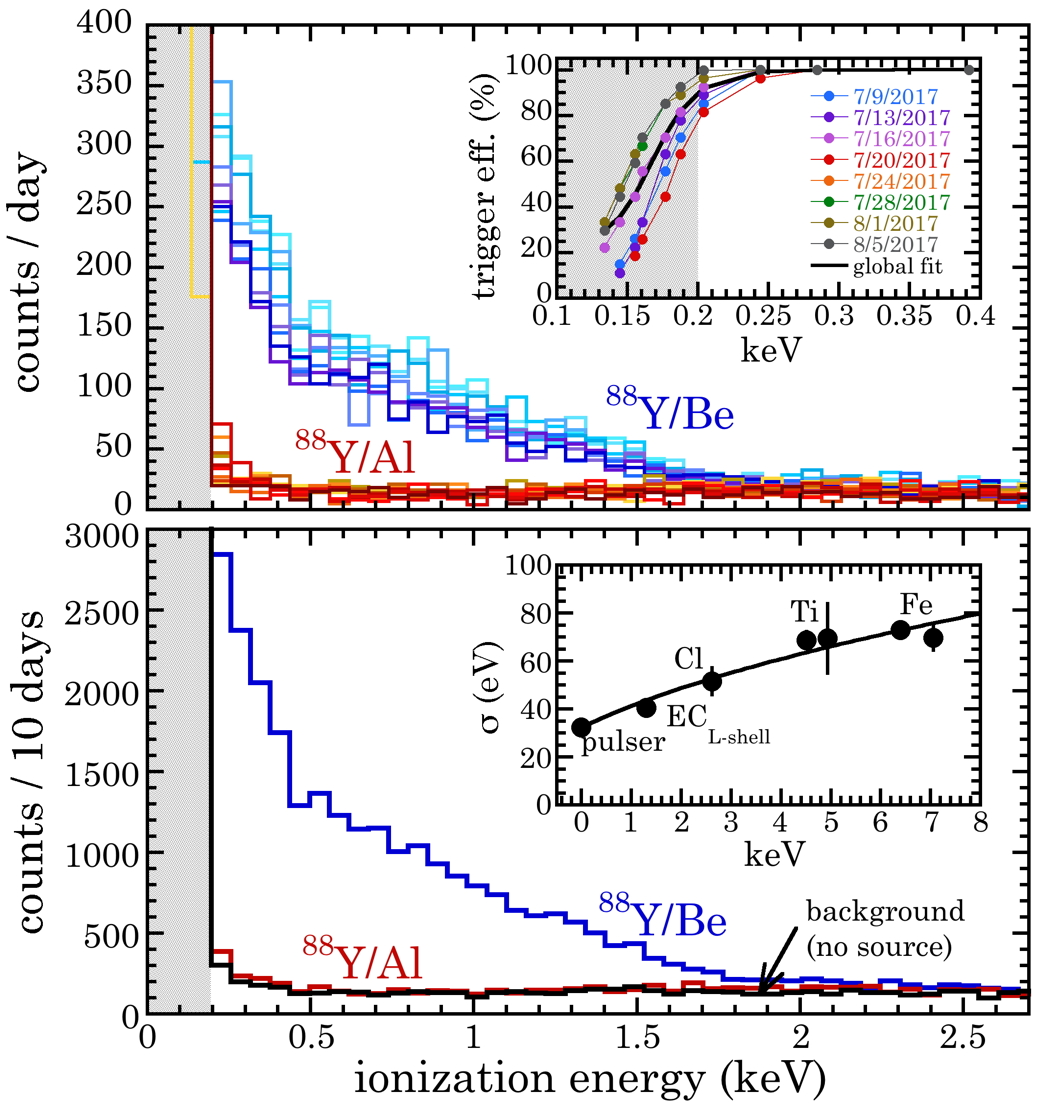}% Here is how to import EPS art
\caption{\label{fig:epsart} $^{88}$Y/Be and $^{88}$Y/Al LEGe spectra for individual daily runs (top) and cumulative (bottom). The grayed region indicates the noise pedestal. The color scale allows to visualize the decay of the source during data-taking. A solid line shows the mean triggering efficiency in the top inset and a fit to the energy resolution, expressed as in \cite{eres1,eres2}, in the bottom inset.}
\end{figure}

The LEGe detector was placed outside of a shield surrounding the source, with 20 cm of lead interposed. Multiple 24 hr exposures alternating both source configurations were taken, with frequent trigger efficiency measurements intercalated, obtained with an electronic pulser. Fluctuations in this efficiency were modest above the 200 eV analysis threshold (Fig.\ 1, inset) and in good correspondence to the specified temperature stability of the digitizer and shaping amplifier used. This last was a  Canberra 2026X-2 customized to provide a 24 $\mu$s shaping time. This results in an optimal  noise performance for detectors with low leakage current. The ionization energy scale and energy resolution were measured using alpha-induced x-ray emission from a number of samples, benefiting from a 25 $\mu$m beryllium entrance window to the LEGe cryostat (Fig.\ 1, inset). 

An immediately evident, reproducible feature in all $^{88}$Y/Be runs is a ``kink" in the spectrum at $\sim$0.5 keV (Fig.\ 1), a peculiarity not reachable during our previous  study with  1 keV ionization threshold \cite{bjornqf}. This rapid rise in low-energy response to NRs came as a surprise: intuitively it is  highly suggestive of a strong deviation from the Lindhard QF model, which is monotonically decreasing with decreasing NR energy \cite{bjornqf}. The data also revealed an excellent  suppression of the gamma component by the lead shield, as per comparison with background in the absence of a source (Fig.\ 1). 

\begin{figure}[!htbp]
\includegraphics[width=.93 \linewidth]{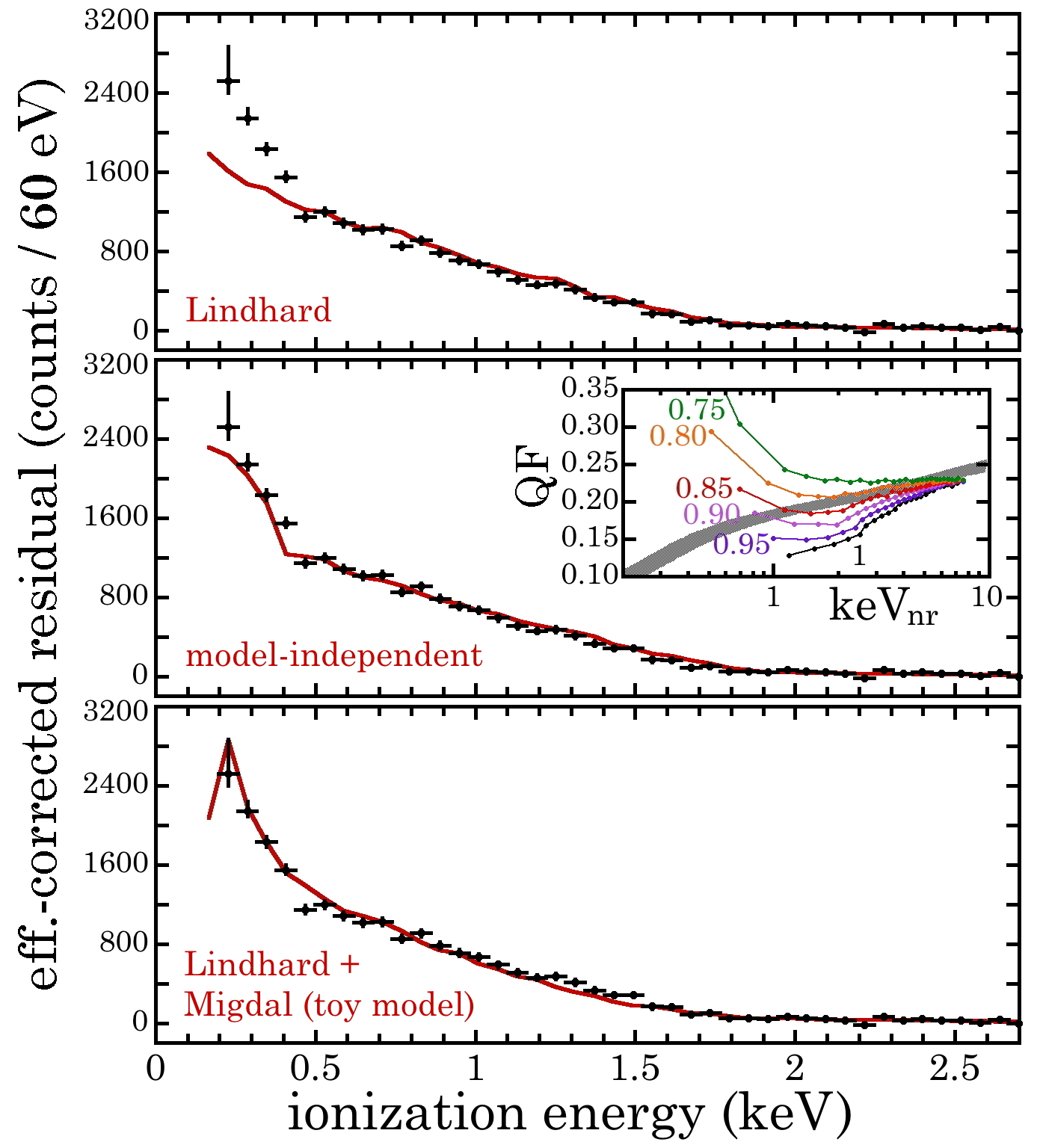}% Here is how to import EPS art
\caption{\label{fig:epsart} $^{88}$Y/Be-$^{88}$Y/Al residual containing NR contributions only. Statistical errors include the effect of triggering efficiency for the  lowest-energy datapoint. Red lines are best-fit simulated models discussed in the text. The inset shows the fractional QF from the model-independent fit vs.\ NR energy. Neutron yields from the source are shown as a fraction $Y$ of its nominal value. A gray band is delimited by two Lindhard fits to data in previous work (Fig.\ 6 in \cite{bjornqf}). }
\end{figure}

Simulations of neutron response were performed using MCNP-PoliMi \cite{polimi}. Attention was paid to including all inner detector and source components in full detail, as well as the effect of impurities in the lead down to ppm level, measured via ICP-MS. The effect of energy resolution was included in the simulations assuming the same Fano factor \cite{fano} applies to both ERs and NRs.  These simulations confirmed the impossibility to explain the low-energy rapid spectral rise in the $^{88}$Y/Be-$^{88}$Y/Al residual while embracing the best-fit Lindhard model presented in \cite{bjornqf}, with or without the presence of the adiabatic factor discussed in that publication. The top panel in Fig.\ 2 shows a direct comparison of the experimental residual with simulated predictions using this Lindhard QF and the nominal average neutron yield of the source. The magnitude of ENDF neutron cross-sections for lead and for materials around the germanium crystal within its cryostat (aluminum and steel) was varied by up to $\pm$50\%. While this  affects the overall rate normalization, it does not give rise to the abrupt spectral feature present below 0.5 keV. 

The absence of a physics-based QF alternative to Lindhard led us to attempt a fit to the residual using a pragmatic, model-independent  approach developed in \cite{alvaro}. This method identifies the observed spectral endpoint of 1.75 keV ionization energy, obtained via a bi-linear fit to the $^{88}$Y/Be-$^{88}$Y/Al residual in the 0.5-3 keV interval, with the maximum NR energy of 8.5 keV$_{nr}$. This initial match is possible only if multiple neutron scattering within the detector is infrequent, as is the case (\mbox{12\%} of simulated neutron histories for this source and geometry). The running integrals of measured interaction rate vs.\ ionization energy and of simulated rate vs.\ NR energy are compared for all energies below their respective endpoints: the dependence of QF on NR energy is inferred from their matching projection onto each other \cite{alvaro}. The neutron yield $Y$ of the source is left as a single free parameter able to provide an optimal fit to the residual. 

This best-fit, capable of reproducing the low-energy excess, is shown as a red line in the middle panel of Fig.\ 2. The corresponding fractional neutron yield of the source is $Y\!=0.86\pm$0.02, with an uncertainty derived from the log-likelihood method used for the fit. The modest disagreement with the nominal average yield ($Y=1$) is representative of our past ability to measure it, as mentioned above. Interestingly, $Y\!\sim\!0.85$ also provides the best match to the Lindhard model above $\sim$2 keV$_{nr}$ up to the 8.5 keV$_{nr}$ endpoint (\mbox{Fig.\ 2} inset). The lowest NR energy that can be explored with this method is \mbox{$Y$-dependent:} however, for all values tested, a trend for a rapid QF increase below $\sim$1 keV$_{nr}$ is noticeable. 

Recent phenomenological work has focused on the Migdal effect \cite{migdal0} and its potential impact on rare event searches \cite{migdal1,migdal2,migdal3,migdal4,migdal5,migdal6}. This phenomenon would account for a prompt emission of excess ionization (``electron shakeoff") following the sudden perturbation to the central atomic potential caused by a NR. While this process has not yet been confirmed for NRs, it has been observed for other  atomic perturbations (e.g, following nuclear $\beta^{\pm}$ decay \cite{shake1,shake2,shake3}). For some detector materials, this excess ionization would significantly increase their sensitivity to low-mass dark matter and CE$\nu$NS  \cite{migdal1,migdal2,migdal3,migdal4,migdal5,migdal6}. The bottom panel in Fig.\ 2 shows that our observations can in principle be understood by invoking a toy model for this process, with free parameters fine-tuned for a good fit. We return to this interesting possibility in Sec.\ VII.

\section{\label{sec:level1}photo-neutron sources: $^{124}$Sb/Be}

A similar procedure was followed using a $^{124}$Sb/Be source. The gamma emitter ($T_{1/2}$=60.2 d) was obtained by activation of a sample of high-purity antimony metal at the North Carolina State University reactor. The average neutron yield during these runs was $\sim 7\times 10^{3}$ n/s. Fig.\ 3 shows the response to the dominant 23 keV  neutron emission from this source \cite{knoll,alvaro}, expected to produce NRs carrying a maximum of 1.3 keV$_{nr}$. A few weeks previous to these tests, an intentional neutron activation of the germanium detector was performed to obtain a 1.3 keV (ionization) energy calibration peak from $^{71}$Ge L-shell electron capture (EC, $T_{1/2}$=11.4 d), part of the bottom inset in Fig.\ 1. A small contamination with the remaining activity under this peak is visible. 

\begin{figure}[!htbp]
\includegraphics[width=.93 \linewidth]{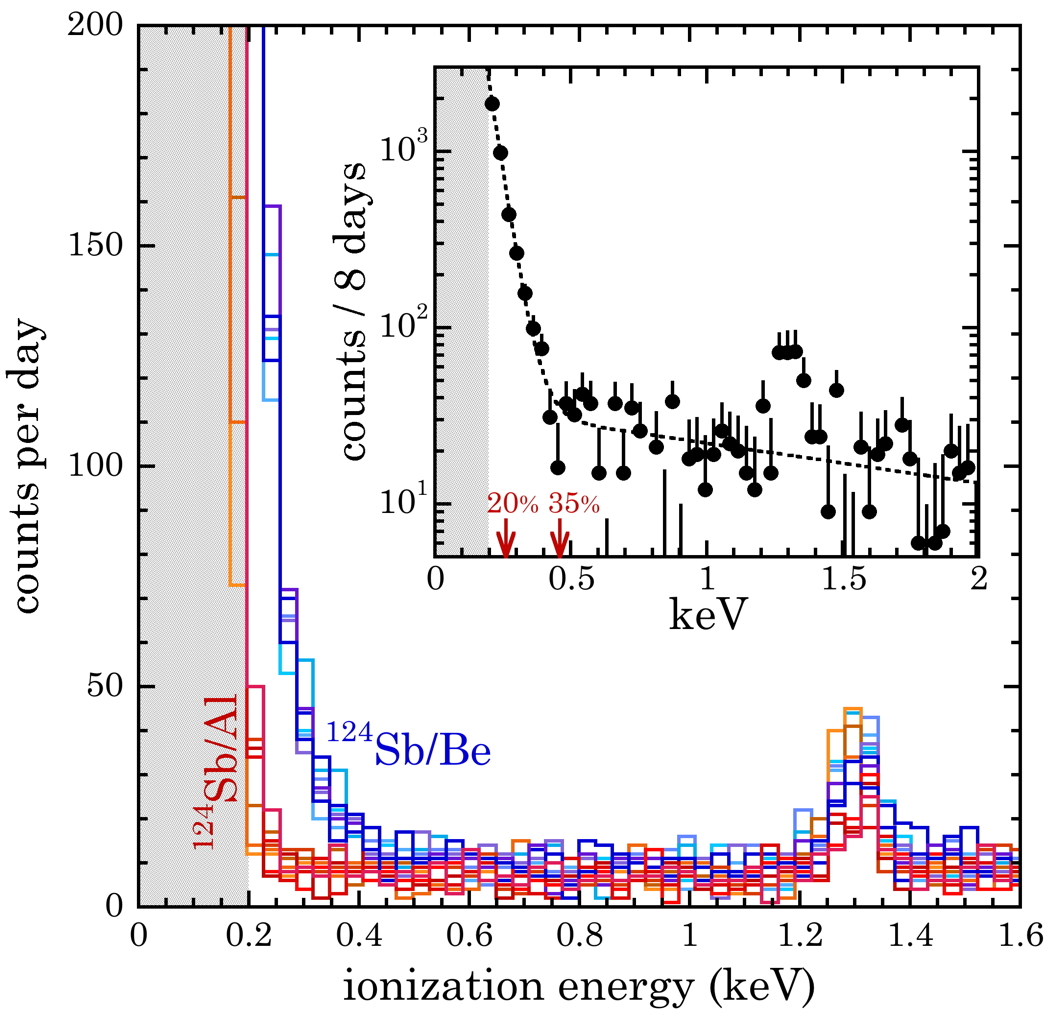}% Here is how to import EPS art
\caption{\label{fig:epsart} LEGe exposure to $^{124}$Sb/Be and $^{124}$Sb/Al. Error bars in their residual (inset, see text) are one-sided for clarity. }
\end{figure}

A full analysis of these data was not attempted due to the small usable energy range spanned by NR signals above the presently-achieved detector threshold and the difficulties in determining a precise ionization endpoint for the NR distribution. Those derive from the presence of a considerable (3\%) branch of higher-energy 378 keV neutrons for this source \cite{branch} and the larger fraction (17\%) of multiple scatters expected from simulations. For comparison, the $^{88}$Y/Be source generates just 0.5\% of neutrons at a higher 963 keV  energy \cite{knoll,ybeprl}. The inset in Fig.\ 3 shows the $^{124}$Sb/Be-$^{124}$Sb/Al residual, fitted by two exponentials representing both neutron branches. Vertical arrows indicate the position of the expected endpoint (1.3 keV$_{nr}$) for single-scatter NRs from the dominant 23 keV branch, for the values of the QF indicated.

\section{\label{sec:level1}Recoils from gamma emission following thermal neutron capture}

While the use of monochromatic low-energy neutrons from photo-neutron sources is a convenient way to produce few-keV$_{nr}$ NRs in the laboratory, this method trusts neutron transport simulations to accurately predict the NR energies being generated. Seeking to  remove this possible source of uncertainty while testing the unexpected results in Sec.\ II, we revisited a technique first put forward in 1975 \cite{jones} able to produce fixed-energy germanium recoils carrying a mere 0.254 keV$_{nr}$, with negligible spread ($\sim$1.5 eV$_{nr}$). These are, to our knowledge, the lowest-energy NRs thus far characterized in any material.

\begin{figure}[!htbp]
\includegraphics[width=.9 \linewidth]{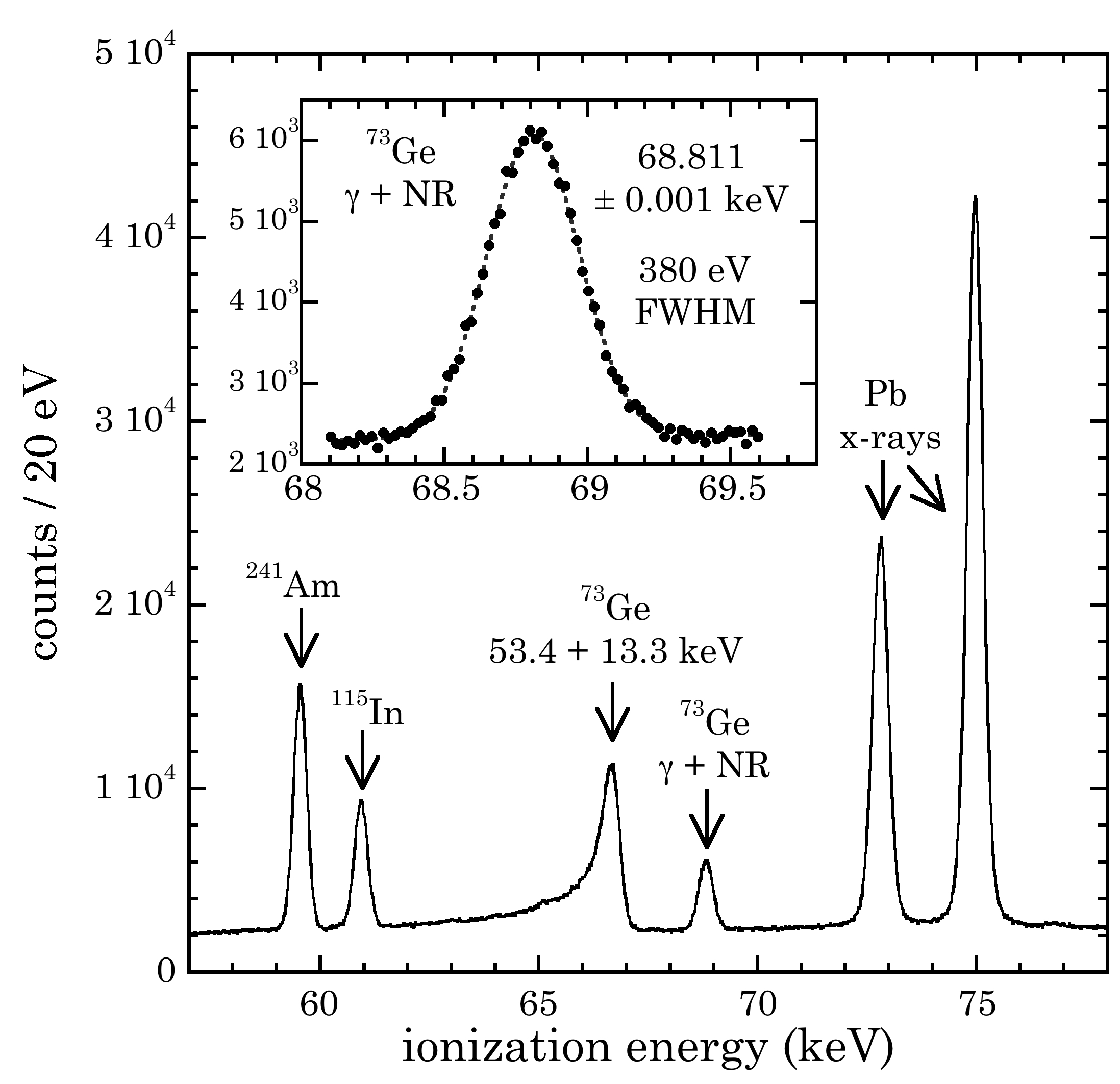}% Here is how to import EPS art
\caption{\label{fig:epsart} Spectrum from a 2.9 hour LEGe exposure to the OSURR thermal beam at 1 kW reactor power, corresponding to $\sim 3\times10^{4}$ n/cm$^{2}$s. The inset expands the ``$\gamma$+NR"  peak from $^{72}$Ge(n,$\gamma$)$^{73}$Ge (see text). As expected from the high purity of this  beam, an excellent Gaussianity is observed. }
\end{figure}

This alternative approach exploits the thermal neutron capture reaction $^{72}$Ge(n,$\gamma$)$^{73}$Ge whenever it populates a 6784.2 keV excited nuclear state. If the reaction takes place within a small germanium detector, short-lived intermediate decays to the lowest  $\sim$68.75 keV $^{73}$Ge excited level  will generate a cascade of high-energy gammas that  escape the crystal with high probability, while inducing the net NR energy listed above. This step is dominated by the emission of a single high-energy gamma \cite{jones}. On the other hand, the $\sim$68.75 keV gamma from the last step in the de-excitation process is  detected with a high efficiency, in combination with the tiny amount of ionization  produced by the recoiling nucleus. For amplifier shaping times sufficiently longer than the lifetime $\tau=0.7~\mu$s of this state both energy depositions are effectively simultaneous. By separately measuring the energy of this gamma alone (e.g., with an auxiliary silicon detector next to an inert  germanium  target exposed to thermal neutrons) it is possible to isolate the ionization energy of the NR and its corresponding QF \cite{jones}.

An ideal beam for this mode of calibration is available at The Ohio State University Research Reactor (OSURR). Single-crystal sapphire and polycrystalline bismuth are used for fast neutron and gamma filtering of core radiations, respectively, resulting in a high-purity thermal neutron beam $\sim$3 cm in diameter with flux selectable in the range $10^{4}$-$10^{7}$ n/cm$^{2}$s. Its cadmium ratio is 266, i.e., there are just 3.76 neutrons with energies higher than 0.4 eV for every 1,000 thermals \cite{ohio}. 

The LEGe detector was placed in the path of this beam. A 7.5\% lithium-polyethylene 1 cm$^{2}$ collimator \cite{shieldw} was used to confine the beam to the region immediately around the germanium crystal, reducing capture gamma backgrounds. As in \cite{jones}, energy calibration peaks were generated continuously during beam irradiation using the dominant gamma emission from a $^{241}$Am source, and lead fluorescence x-rays from a 1 mm Pb disk placed blocking the LEGe Be window, while holding a $^{57}$Co source (Fig.\ 4). To avoid the introduction of any bias in the analysis, the same nominal energies as in \cite{jones} were assigned to these calibration peaks. All peaks in the vicinity of the sought signal ($^{73}$Ge ``$\gamma$+NR") are shown in Fig.\ 4. A skewed peak at $\sim$66.7 keV arises from the effect of using a 8 $\mu$s amplifier shaping time on a $^{73}$Ge  de-excitation cascade involving the emission of a 53.4 keV gamma followed by 13.3 keV from a level with a half-life of 2.9 $\mu$s. A peak at \mbox{60.916 keV} \cite{iaea} originates from  activation of an indium electric contact on the surface of the crystal. 

Fig.\ 5 compares the position of the sought $\gamma$+NR  peak with that previously obtained in \cite{jones}. The presently achieved uncertainty is considerably smaller than in the original 1975 measurement. Our result is also incompatible with it, pointing in the direction of a larger QF. Prompted by this observation, an attempt to independently measure the energy of the isolated gamma was made. During a second OSURR visit, a 0.5 g sample of $>96$\% isotopically-enriched $^{72}$Ge oxide \cite{ge72} was exposed to the collimated beam, using an Amptek XR-100SDD silicon x-ray detector in close proximity to collect the gamma emissions from this target. To avoid an excessive activation of the detector itself by neutrons scattered from the irradiated sample, a 3 mm-thick blanket of $^{6}$LiF powder (95\% isotopically-enriched in $^{6}$Li \cite{aldrich}) compacted to a density of 1.1 g/cm$^{3}$ was inserted between both. The powder was held in place by aluminum foils glued to a supporting acrylic ring.  This blanket is $\sim$95\% transparent to 68.75 keV gammas, while it reduces thermal neutron transmission by close to four orders of magnitude. A gamma-less neutron capture via $^{6}$Li(n,t)$^{4}$He avoids a background penalty. Unfortunately, even with this precaution, several peaks traceable to activation of tungsten in a multilayer collimator internal to this  silicon detector  encumbered the sought signal.

\begin{figure}[!htbp]
\includegraphics[width=.8 \linewidth]{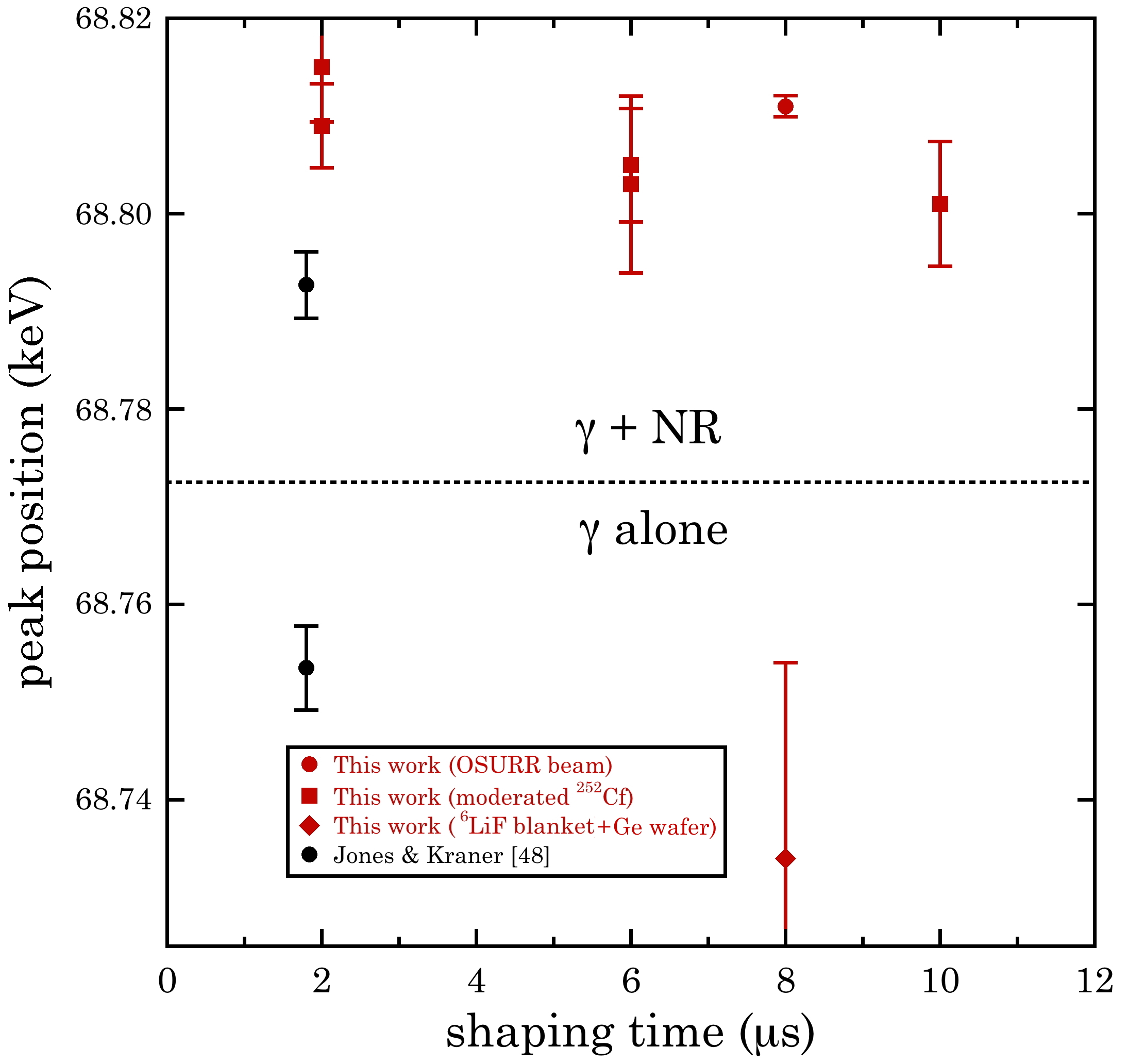}% Here is how to import EPS art
\caption{\label{fig:epsart} Comparison of $^{72}$Ge(n,$\gamma$)$^{73}$Ge  measurements containing simultaneous gamma and NR energy depositions, with those involving the gamma energy alone, as a function of amplifier shaping time.  Their difference corresponds to the ionization energy deposited by  0.254 keV$_{nr}$ NRs in germanium \cite{jones}. The same data acquisition system (amplifier, multi-channel analyzer) was used for all present measurements. }
\end{figure}

An alternative route to this ancillary measurement, finally successful, is depicted in Fig.\ 6. Thermal neutrons from moderated  $^{252}$Cf  activate a germanium metal sample consisting of two wafers adding up to 1 mm thickness, placed in  proximity to the LEGe detector Be window. The thermal flux reaching the LEGe crystal is abated by the same $^{6}$LiF blanket as above, in addition to a cadmium metal sheath surrounding the cryostat. The residual between spectra obtained with and without the presence of the germanium wafers, shown in Fig.\ 7, reveals a peak at the position of the ``$\gamma$ alone" emissions from this sample, with expected width (341$\pm$49 eV FWHM, compare with Fig.\ 4). The energy scale was continuously defined as above, assisted by $^{241}$Am and Pb fluorescence peaks. Our ``$\gamma$ alone" peak position is seen to be compatible within errors with the original 1975 measurement (Fig.\ 5). The present uncertainty was limited by the weak intensity ($4\times10^{5}$ n/s) of the available  $^{252}$Cf source. A discussion contained in \cite{jones} points at an even better agreement with our  peak position when  their  $^{73}$Ge($\alpha$,$\alpha$'$\gamma$)$^{73}$Ge ``$\gamma$ alone" datapoints \cite{jones2}, identified in \cite{jones} as outliers, are dropped. 

\begin{figure}[!htbp]
\includegraphics[width=.8 \linewidth]{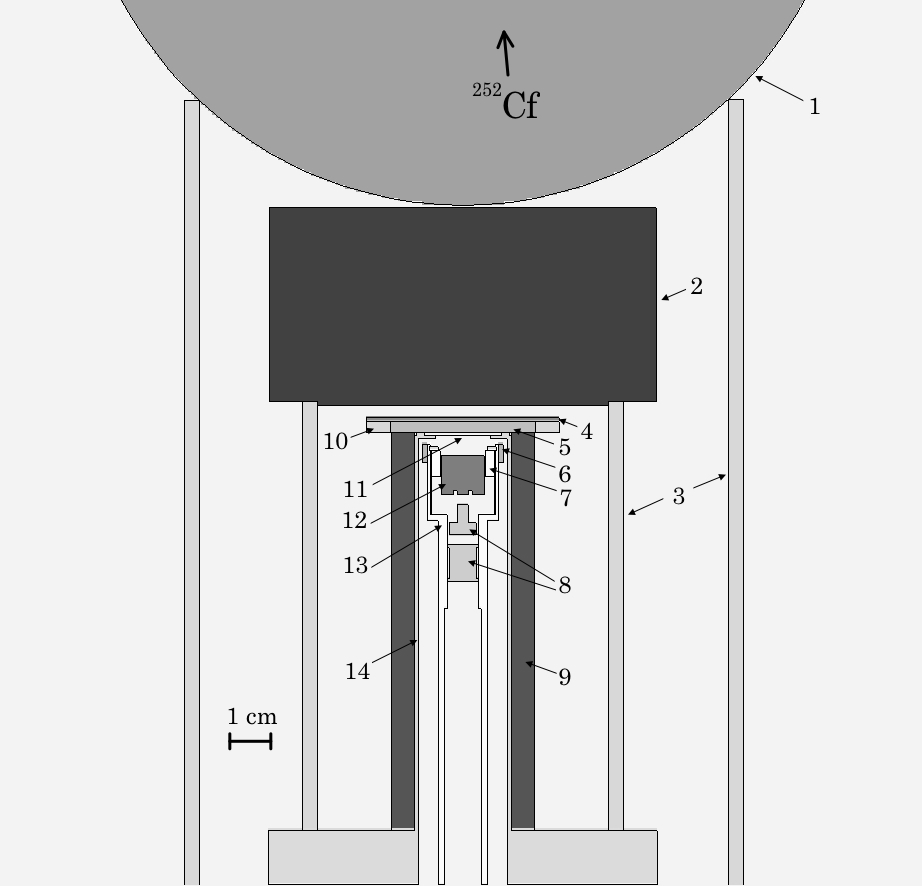}% Here is how to import EPS art
\caption{\label{fig:epsart} Geometry employed for the ``$\gamma$ alone" measurement described in the text: 1) 20.3 cm-diameter polyethylene sphere housing a $^{252}$Cf source at its center, 2) Pb disc (fission gamma shield, fluorescence source), 3) cylindrical acrylic holders, 4) Ge wafers, 5) $^{6}$LiF blanket, 6) polycarbonate ring, 7) Al crystal holder, 8) PTFE field-effect transistor holder, 9) layered Cd sheath, 10) acrylic ring, 11) Be window, 12) Ge crystal, 13) Al cold finger, 14) stainless steel cryostat. A  $^{241}$Am source was positioned near the crystal, outside the Cd.  }
\end{figure}

\begin{figure}[!htbp]
\includegraphics[width=.9 \linewidth]{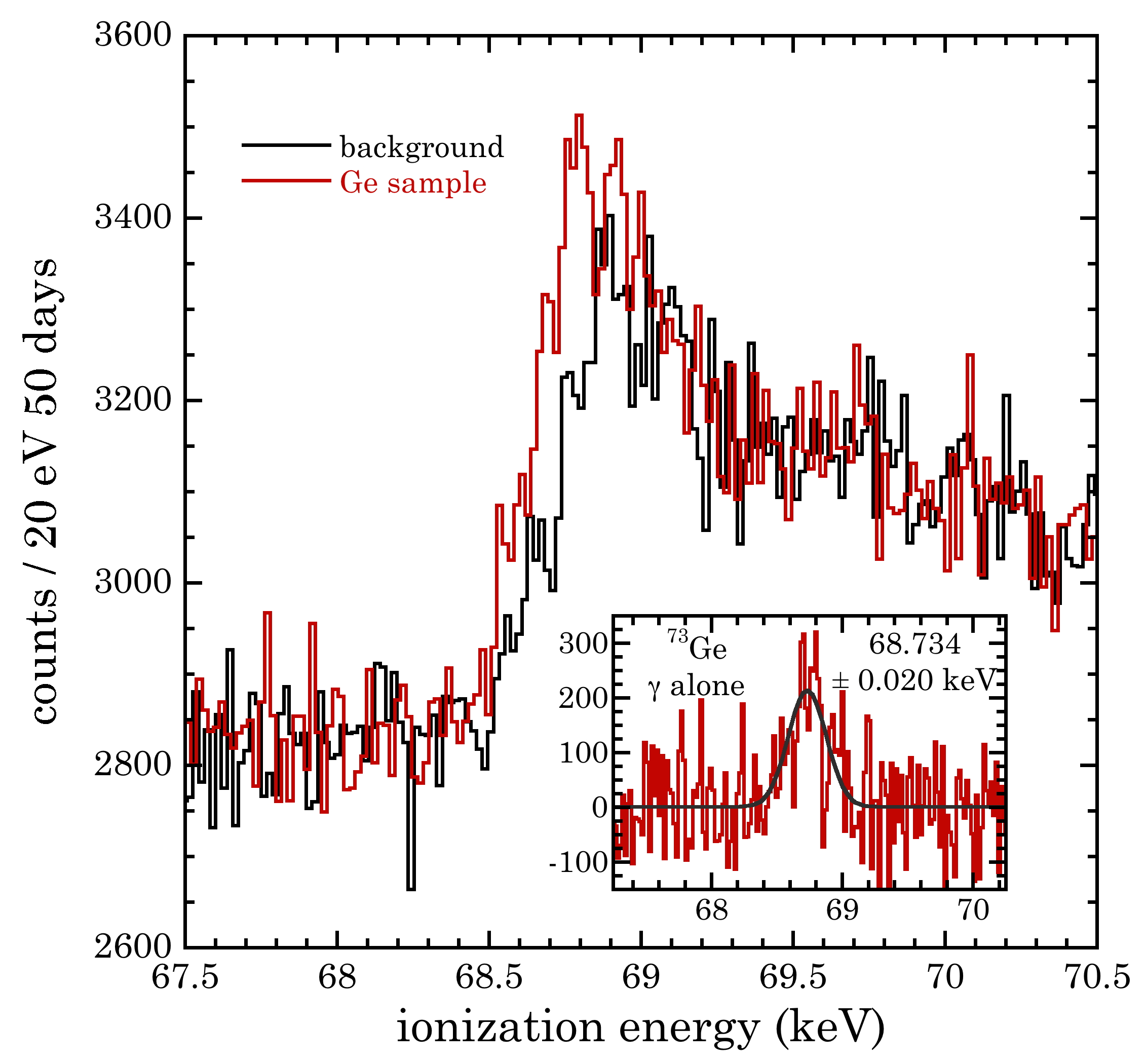}% Here is how to import EPS art
\caption{\label{fig:epsart} Normalized spectra obtained using the arrangement of Fig.\ 6, in presence and absence of Ge sample (wafers). Their residual and Gaussian fit are shown in the inset. The ``shark tooth" structure above $\sim$68.5 keV originates in LEGe capture of intermediate neutrons able to traverse Cd sheath and $^{6}$LiF blanket \cite{shark1,shark2}. Neighbouring $^{241}$Am and Pb x-ray energy calibration peaks fall outside the  range of this figure. }
\end{figure}

We considered the possibility that the $\sim$1.8 $\mu$s shaping time used in \cite{jones} (based on the 4 $\mu$s peaking time quoted) might have been insufficient to account for the  $\tau=0.7~\mu$ lifetime of the 68.75 keV $^{73}$Ge excited level, shifting the position of their $\gamma$+NR peak to a lower value. This hypothesis was tested by removing Ge wafers, $^{6}$LiF blanket, and cadmium sheath from the configuration in Fig.\ 6, allowing the LEGe to capture thermalized $^{252}$Cf neutrons unimpeded, while varying the shaping time of our amplifier. %The removal of the $^{241}$Am source was noticed to have an advantageous effect on background in this configuration. Its calibration reference was replaced by the $^{115}$In peak in Fig.\ 4, the position of which is known to 1 eV precision \cite{iaea}). 
As can be seen in Fig.\ 5 (``moderated $^{252}$Cf"), no significant dependence on shaping time can be concluded. However, the mean of these $\gamma$+NR measurements (68.808$\pm$0.0025 keV) is in agreement with our  OSURR datapoint (68.811$\pm$0.001 keV), reinforcing the tension with the previous equivalent in \cite{jones} \mbox{(68.793$\pm$0.0034 keV).} The much higher statistics under $\gamma$+NR and energy calibration peaks in  OSURR data, compared to those in \cite{jones}, should be emphasized at this point.

All in all, our present revival of the thermal capture technique first described in \cite{jones} points at 77$\pm$20 eV of ionization being produced by a 0.254 keV$_{nr}$ germanium NR, corresponding to a 30.3$\pm$7.9\% QF at this recoil energy. This is in clear contrast with the 15.4$\pm$2.1\% QF arrived at in \cite{jones}. Enticingly, as noticeable in the inset of Fig.\ 2, this additional QF measurement is nevertheless  compatible with an extrapolation of the $^{88}$Y/Be model-independent best-fit ($Y$=0.86) to  lower recoil energies.

\section{\label{sec:level1}monochromatic 24 keV iron-filtered neutrons}

In view of the apparent consistency of our thermal neutron capture and photo-neutron results, we embarked on a final characterization effort able to provide additional QF values at discrete recoil energies below 1 keV$_{nr}$. To that effect, the LEGe was exposed to a highly-monochromatic 24 keV ($\pm$2 keV FWHM) iron-filtered neutron beam available at the Kansas State University (KSU) experimental reactor. Its design, construction, and characterization are described in \cite{ksu}. We previously used this filter to measure the germanium QF in the range $\sim$0.7-1.2 keV$_{nr}$ \cite{ppc}, however using a large detector ($\times$90 the LEGe volume) prone to multiple scattering, with  $\sim$2.5 times the intrinsic electronic noise of the present device. In these conditions, the extraction of QF values at such low energies required a  complex data analysis \cite{phil}, in contrast with the straightforward approach presently possible. 

\begin{figure}[!htbp]
\includegraphics[width=.8 \linewidth]{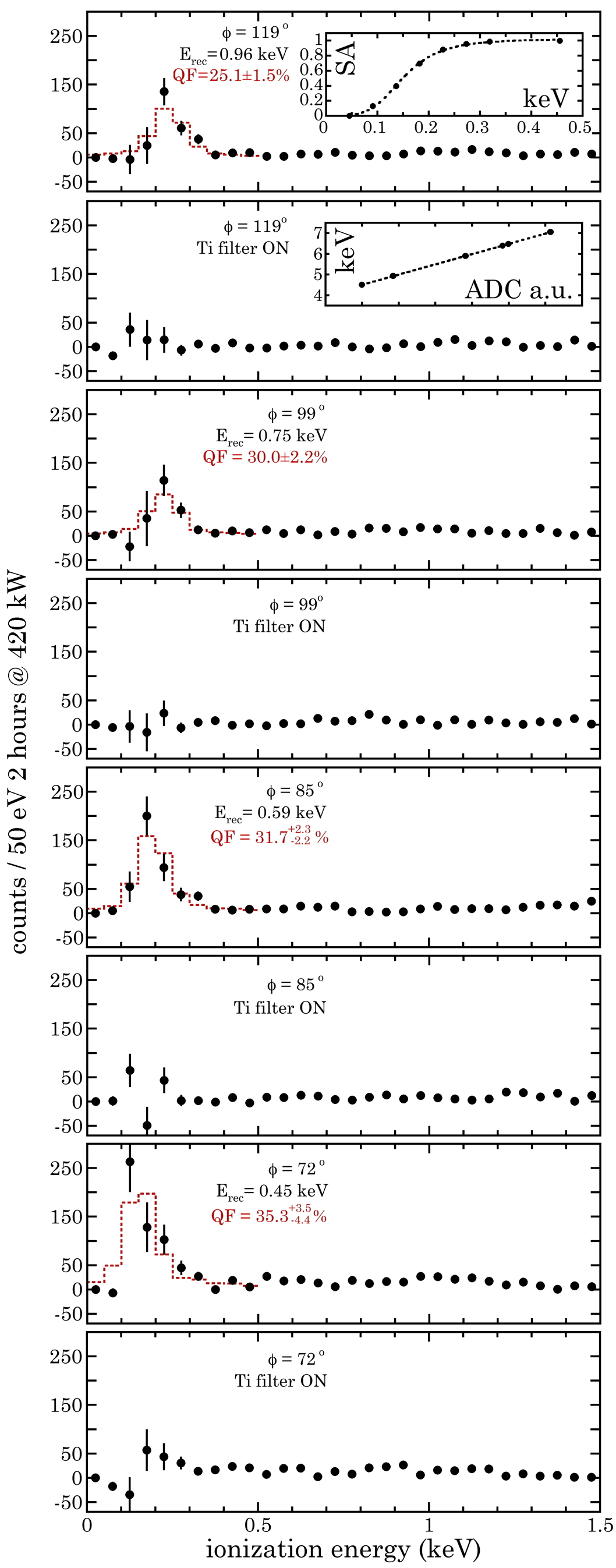}% Here is how to import EPS art
\caption{\label{fig:epsart} Normalized residual spectra for all KSU iron-filter runs, labelled by neutron scattering angle $\phi$, nominal NR energy,  and Ti-filter on/off status. Signals produced by NRs from 24 keV neutron scattering are visible only in Ti-filter absence.  Red histograms show a simulated response, for the best-fit QF listed. A small  increase in signal rate with decreasing $\phi$ is expected from scattering kinematics. Error bars in the insets are encumbered by data point size.}
\end{figure}

This technique exploits a dip in the elastic scattering neutron cross-section for iron at 24 keV, where it is reduced by three orders of magnitude over a narrow energy region. For ultra-pure iron filters $\sim$1 m in length  carefully designed to reduce capture gamma backgrounds, a high beam purity is achievable \cite{ksu}. A convenient feature of the method is the ability to ``switch off" the dominant 24 keV beam component while preserving all beam contaminations, by inserting an additional thin (1.25 cm) titanium post-filter \cite{ksu}. This exploits a resonance in the Ti cross-section precisely at $\sim$24 keV neutron energy. A comparison of Ti filter on/off runs can provide convincing evidence that low-energy  signals assigned to NRs do indeed originate from the scattering of 24 keV neutrons. 

Data were acquired using the system in \cite{bjornqf}, able to digitize LEGe preamplifier and backing detector signals at 120 MS/s. A trigger was provided by a single-channel analyzer centered around the neutron capture peak of a large (1.5 cm $\times$ 5 cm diam.) 95\% isotopically-enriched $^{6}$LiI[Eu] scintillator, used to detect neutrons scattered off the LEGe. The virtues of this choice of backing detector for this application are discussed in \cite{ksu}. The scintillator was mounted on a support arm able to pivot around the LEGe cryostat, 15 cm away from the germanium crystal. A goniometric table installed on the LEGe Dewar allowed the selection of scattering angles and their corresponding NR energy. A best effort was made to center the small germanium crystal in the Gaussian-profile ($\sim$5.9 cm FWHM \cite{ksu}) neutron beam, using a system of two alignment lasers.  A perfect alignment could not be guaranteed due to the large ($\sim$1.1 m) detector distance to the beam exit, imposed by the  shielding necessary to block capture gammas from the reactor wall. Nevertheless, the observed NR signal rates  agreed within a factor of two with simulated predictions based on beam characterization data \cite{ksu}, for the four scattering angles tested. 

Measurements were performed over two days of beam availability, with roughly two-hour runs for each scattering angle and Ti-filter on/off configuration. Reactor poisoning forced to progressively decrease its power from 420 kW to 300 kW. The results presented here are normalized to the same exposure and power for all runs. The energy calibration of the LEGe detector was performed immediately before and after the completion of all beam runs, using the alpha-induced x-rays from Ti and Fe samples (Fig.\ 1) and a $^{55}$Fe source, for a total of six reference peaks in the 4-7 keV energy range (Fig.\ 8 inset). No measurable energy drift was observed over the two days of data-taking: all twelve calibration points are used for the linear fit in the figure. 

During data analysis, events in coincidence and in anti-coincidence with the neutron capture signal from the backing detector were inspected using an edge-finding algorithm, illustrated using this same LEGe detector in \cite{muon}. It is able to identify the rising edge characteristic of radiation-induced pulses in  preamplifier traces, while rejecting low-energy noise nuisances. The signal acceptance (SA) of this algorithm was measured (Fig.\ 8 inset) using programmable electronic pulser signals of same rise-time as calibration x-rays. The energy of events passing this data-quality cut was determined using a digital implementation of a 36 $\mu$s zero-area cusp filter \cite{zac1,zac2,zac3}. The residual difference of  energy spectra from coincident and anti-coincident events, normalized to same exposure, is expected to be dominated by the NR signals sought and free of  significant low-energy noise by virtue of the subtraction. These residuals, corrected for SA, are shown in Fig.\ 8. Error bars are statistical: by making the anti-coincidence time window a factor of fifty longer than that for coincidence, the uncertainty on steady-state low-energy backgrounds is greatly reduced. As expected, clear low-energy excesses are observed only in the absence of the Ti-filter, confirming their origin in the elastic scattering of 24 keV neutrons. 

The distribution of MCNP-PoliMi simulated energy depositions by NRs in the germanium crystal  also producing a capture in the backing detector was converted into an ionization equivalent using QF values in the range 15\%-40\%, including the effect of energy resolution and multiple scattering (4.1\% to 7.6\% of all events), for each of the four scattering angles tested. The resulting spectra were compared to the Ti-off residuals of Fig.\ 8, obtaining best-fit QF values  via  log-likelihood analysis. The QF errors shown in Figs.\ 8 and 9 combine the uncertainty extracted from this procedure with that from the energy calibration. The energy spread (HWHM) of simulated events is utilized as horizontal error bars in Fig.\ 9. 

\begin{figure}[!htbp]
\includegraphics[width=.87 \linewidth]{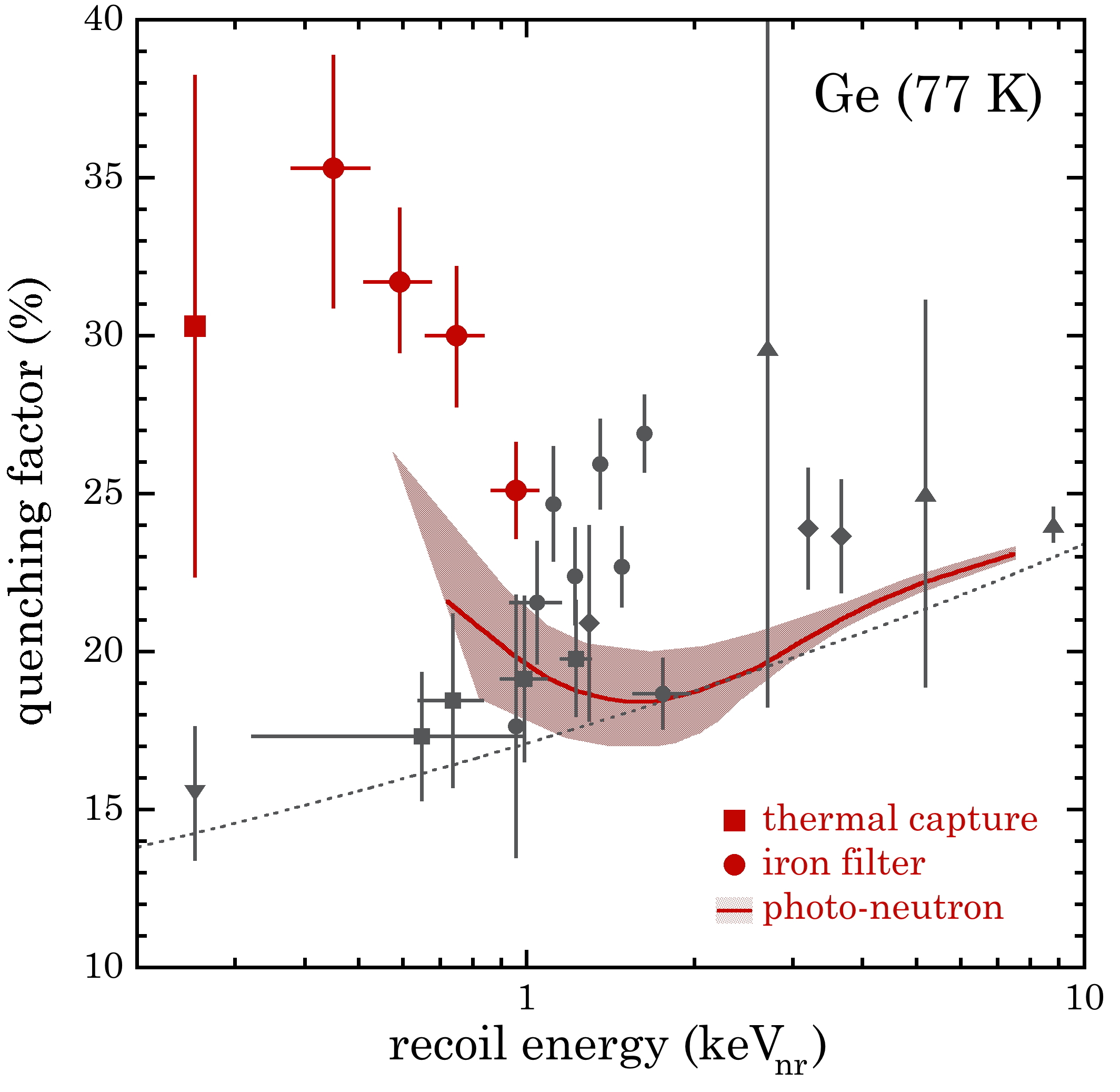}% Here is how to import EPS art
\caption{\label{fig:epsart} Present QF results, labelled by calibration technique. A red band shows the 95\% C.L. region for the model-independent fit of Fig.\ 2. A dotted line is the Lindhard model with a default germanium value of $k$=0.157 \cite{bjornqf}. Previous measurements are shown in gray: circles \cite{jones2}, squares \cite{ppc,phil}, diamonds \cite{texono2}, triangles \cite{messous}, and inverted triangle \cite{jones}.}
\end{figure}

\section{\label{sec:level1}Systematics and Compatibility}

In this section we elaborate on possible systematic effects able to have a moderate impact on our  measurements, as well as on the compatibility of these measurements with each other and with previous work in \cite{bjornqf}.

The normalization of the two datasets shown in \mbox{Fig.\ 7} is based on a matching of their backgrounds in the regions above and below the peak of interest. The same normalization factor of 1.51 was found to apply to both regions. An alternative method of normalization based strictly on the difference in exposure between the datasets would use a factor of 1.57. This modest difference is due to the small yet finite shielding of backgrounds that the thin germanium sample produces. If this alternative normalization is employed, the ``$\gamma$ alone" peak position is minimally shifted by 11 eV to a lower energy. The net result is an increase in the QF derived from the thermal capture method to 34.6$\pm$7.9\%, improving the compatibility of this data point with the trend of those derived from the KSU iron filter (Fig.\ 9). 

The independent term in the linear fit used to correlate x-ray energy to analog-to-digital converter (ADC) amplitude units (Fig.\ 8, inset) has a finite value of \mbox{49 eV}, with a small uncertainty of $\pm$11 eV due to the excellent linearity observed (Pearson's R=0.99998). Use of lower-energy calibration data points  (e.g., via alpha irradiation of PVC, producing Cl x-rays as in Fig.\ 1) was not possible due to the longer exposures required for those and the limited beam time available at KSU. The quality of this fit is not in doubt, as non-zero independent terms of this magnitude are commonplace and traceable to the algorithms used for energy determination. It is however worth mentioning that making this independent term equal to zero would bring the iron-filter data points in Fig.\ 9 to near-perfect agreement with the photo-neutron best fit shown there and its extrapolation to low energy. Nevertheless, as mentioned in Sec.\ V, the effect of the known uncertainty in the energy scale is already included in the vertical error bars for iron-filter data. 

The photo-neutron QF should be considered an approximation, as its model-independent method is predicated  on a total absence of multiple scatters. Still, a  more complex analysis leaving both source yield $Y$ and ionization endpoint in the $^{88}$Y/Be-$^{88}$Y/Al residual as free parameters might be able to produce an improved fit to the low-energy excess in Fig.\ 2, bringing the derived QF closer to iron-filter results. In lieu of this analysis, we assess the agreement between both techniques by assuming a QF model consisting of a no-frills linear fit to the iron-filter QF datapoints in Fig.\ 9 for energies below 1.35 keV$_{nr}$. At this energy this fit intersects the standard $k$=0.157 Lindhard line in the same figure. For higher energies the QF model switches to Lindhard. When this simple model is applied to the interpretation of photo-neutron data, a fair quantitative and qualitative agreement is obtained (Fig.\ 10). This test illustrates the internal consistency of the ensemble of our measurements. 

\begin{figure}[!htbp]
\includegraphics[width=.9 \linewidth]{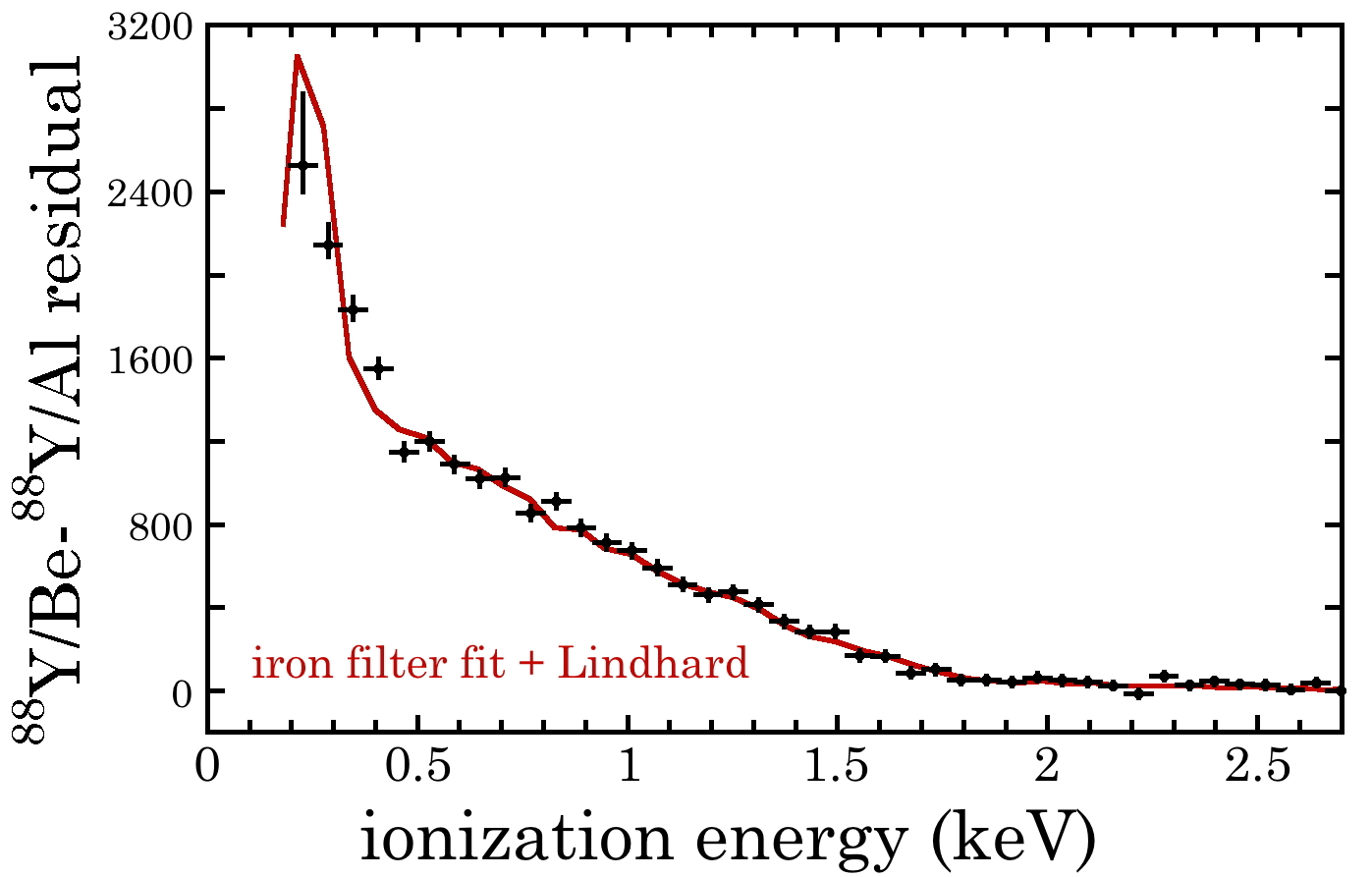}% Here is how to import EPS art
\caption{\label{fig:epsart} Comparison of the  QF model described in the text, based on iron-filter QF measurements and Lindhard theory, with the $^{88}$Y/Be-$^{88}$Y/Al residual of Sec.\ II. The adopted source yield is $Y=0.95$. No other free parameters are used.}
\end{figure}

Finally, we have examined the compatibility of the new measurements presented here with our previous photo-neutron dataset in \cite{bjornqf}. As mentioned in Sec.\ I, the detector used for that study had a threshold five times larger than presently achieved. This derived from a combination of higher intrinsic electronic noise and an issue with the internal gain of the digitizer employed, later resolved. As a result, signals from NRs below $\sim$4 keV$_{nr}$ could only be detected as part of events involving multiple neutron scattering, dominant for that large crystal. In addition to this, two free parameters had to be included in the analysis to account for the thickness of dead and transition surface layers \cite{deadl} in that p-type diode. Added to the yield of the source and the two free parameters allowed for the Lindhard model (the mentioned $k$ and one to account for a possible adiabatic factor \cite{ahlen1}), this resulted in a total of five free parameters being used to match simulation to data.

Two QF models based on present experimentation, both devoid of free parameters,  have been tested against the $^{88}$Y/Be-$^{88}$Y/Al residual from  \cite{bjornqf}. As in that publication,  we used a popular Markov Chain Monte Carlo (MCMC) ensemble sampler \cite{mcmc1,mcmc2} to explore the  parameter space available to the fits. The first model is that described in the discussion above regarding Fig.\ 10. The second corresponds to the photo-neutron best-fit line in Fig.\ 9, linearly extrapolated to lower energy. Three free parameters were adopted, the yield of the source $Y$, and the independent thicknesses of dead and transition layers. 

As a first cross-check, we reproduced the results in \cite{bjornqf} for the Lindhard model, finding similar best-fit values for all five free parameters. However, when using the new QF models, the quality of their fits to the  residual is comparable to that using Lindhard. What is more, both present QF models result in a combined depth for surface layers $\sim$40\% shallower than in \cite{bjornqf}. We consider this to be more reasonable, as the values obtained in \cite{bjornqf} would have resulted in a sizable degradation of the energy resolution, not observed. The new models also favor $Y\sim$1.18, in considerably better agreement with source characterization than the previous $Y$=1.37 \cite{bjornqf}. We conclude that while the dataset in \cite{bjornqf} was sufficient to constrain  free parameters in a model  (Lindhard) embraced as an article of faith, it is inadequate to exclude QF deviations  taking place at energies well-below detector threshold. This is in contrast with the model discrimination possible with the current detector and dataset, illustrated in Fig.\ 2, while using an economy of free parameters (one). In all, we consider that the present effort supersedes that in \cite{bjornqf}.

\section{\label{sec:level1}Commentary and conclusions}

Our measurements strongly suggest that a new physical process (or processes), absent from Lindhard's classical treatment of ion slowdown, dominates the production of ionization by NRs below $\sim$1.3 keV$_{nr}$ in a germanium semiconductor drastically enhancing the low-energy quenching factor of this material. Unexpected as this may seem, a similar behavior has been observed before for the production of light by sub-keV proton recoils in organic scintillators \cite{ahlen1,ahlen2}. This property was recently confirmed \cite{ej301}.  In this section we briefly comment on possible origins for our observations, and on their implications for upcoming rare-event searches. 

A plausible explanation for the observed behavior is the Migdal effect already invoked in Sec.\ II.  The toy model adopted for this process, shown in \mbox{Fig.\ 2}, assumes that Migdal-style electron shakeoff takes place for a significant fraction of NR episodes. A shakeoff probability $P=$ 50\%  is a first free parameter fine-tuned to obtain the agreement with data shown in the figure. A second free parameter $\epsilon=$ 0.35 keV describes the kinetic energy $E$ of the  single electron ejected, distributed as $\propto\!e^{-E/\epsilon}$ up to a maximum constrained by the magnitude of the atomic perturbation (simulated initial NR energy) and conservation of energy. This functional form used to sample $E$ is a very crude approximation to formal Migdal differential ionization probabilities in \cite{migdal1}. Following this same publication, the ejected electron is assumed to be preferentially originating from the germanium M-shell. The energy invested in breaching an electron binding energy of 35 eV is taken to be returned as ionization, following atomic orbital relaxation via radiative or Auger emissions. The remaining energy up to that of the simulated initial NR is dispersed assuming that a Lindhard QF as in \cite{bjornqf} applies to the slowing-down of the recoiling ion. NRs not producing shakeoff are Lindhard-governed. 

The chosen values of $P$ and $\epsilon$ can be further adjusted to obtain similar fits to photo-neutron data. The nominal $Y=$ 1 neutron yield from the source was adopted for the fit shown in  Fig.\ 2. A value of $P=$ 50\% may seem arbitrarily large, being a factor of approximately seven above the integrated ionization probabilities calculated for Migdal shakeoff from atomic germanium in \cite{migdal1}. However, recent work  \cite{migdal4} indicates that this probability is significantly enhanced for the present case of a germanium semiconductor. It is nevertheless hard to quantify the magnitude of this probability increase using the  information  in \cite{migdal4}, specific to low-mass WIMPs.  Present data offer a benchmark against which  phenomenological Migdal predictions can be contrasted, possibly confirming the presence of this process. 

Alternative paths leading to an enhanced ionization yield from low-energy NRs in semiconductors have been recently put forward in  \cite{yoni1,yoni2,yoni3}.  
%Statistical fluctuations in the number of information carriers (electron-hole pairs here) generated by  radiation interactions may also be considerably larger for low-energy NRs than for their ER counterparts \cite{lfano1,lfano2,lfano3,lfano4}. 
All these interpretations rapidly complicate  when the secondary NRs abundantly  produced in the wake of a low-energy primary NR are folded in. As a reference, a 1 keV$_{nr}$ germanium recoil produces on average 43 displaced secondaries, each carrying just a few tens of eV$_{nr}$, adding up to 92\% of the   primary energy, with essentially every atom within a (25\AA)$^{3}$ lattice voxel containing the full  trajectory of the primary being perturbed \cite{srim}. From this perspective, focus on the primary as in our Migdal toy model should be replaced by an  understanding  of how a densely-packed cloud of secondaries might collectively or individually contribute to a higher ionization yield. Examined from this point of view,  differences between crystals in their spatial concentration of  lattice excitation at the NR site may become very relevant: we  observed a {\it reduction} in  QF with respect to  Lindhard  for a lighter, lower-density semiconductor, silicon, at least down to 0.7 keV$_{nr}$ \cite{alvaro}. A possible contrast between germanium and silicon in their response to sub-keV NRs is an incipient   area of study \cite{cutoff}.

The importance of developing material-specific models of response to sub-keV NRs, solidly anchored on experimental characterization data, cannot be overemphasized. If our observations are confirmed, ionization-sensitive germanium detectors with thresholds below $\sim$200 eV \cite{ESS,cdex,supercdms} should enjoy a sizeable improvement in their sensitivity to low-mass WIMPs and to CE$\nu$NS signals from reactor antineutrinos (for a 300 eV threshold \cite{conus2} our observations have a comparatively minor impact). While this may be advantageous in the second context when such a detector is simply used as a neutrino counter for reactor monitoring \cite{adam,huber}, an imperfect understanding of the sub-keV QF would severely hamper the opportunities for probing new physics that CE$\nu$NS otherwise affords \cite{csiqf}. Aware of both promise and perils, we conclude by encouraging further phenomenological predictions in this low-energy frontier, and welcoming innovative experimental techniques capable of their verification  \cite{french}.

\begin{acknowledgments}
We are indebted to Dan Baxter, Yoni Kahn and Gordan Krnjaic  for calling our attention to the Migdal effect as a possible origin for our observations and for many useful conversations on the subject. Similarly, to Alvaro Chavarria for proposing the use of the model-independent method for $^{88}$Y/Be analysis, and to  Simon Knapen and Tongyan Lin for helpful input.  Our gratitude also goes to Jim Colaresi and Mike Yocum at Canberra for  detector-related consultations. Three experimental reactors were involved in this work: we thank Lei Cao, Andrew Kauffman, and Susan White at Ohio State University, Scott Lassell at North Carolina State University, and Alan Cebula at Kansas State University for their generous support of our operations at their facilities. This work was funded by NSF awards PHY-1806722 and PHY-1812702, and by the Kavli Institute for Cosmological Physics at the University of Chicago through an endowment from the Kavli Foundation and its founder Fred Kavli. 
\end{acknowledgments}

%\appendix
%\section{\label{sec:level1}Systematic uncertainties}

\bibliography{apssamp}% Produces the bibliography via BibTeX.

\end{document}